\documentclass{aa}
\usepackage{txfonts,graphicx,natbib}
\bibpunct{(}{)}{;}{a}{}{,}

\begin{document}

\title{Multicolor observations of the afterglow of the short/hard GRB\,050724%
\thanks{Based on observations carried out at ESO telescopes under programmes Id
075.D-0787, 075.D-0468 and 078.D-0809.}}

\author{
D. Malesani\inst{1} \and
S. Covino\inst{2} \and
P. D'Avanzo\inst{2,3} \and
V. D'Elia\inst{4} \and
D. Fugazza\inst{2,5} \and
S. Piranomonte\inst{4} \and
L. Ballo\inst{6} \and
S. Campana\inst{2} \and
L. Stella\inst{4} \and
G. Tagliaferri\inst{2} \and
L. A. Antonelli\inst{4,7} \and
G. Chincarini\inst{2,5} \and
M. Della Valle\inst{8,9} \and
P. Goldoni\inst{10,11} \and
C. Guidorzi\inst{2,5} \and
G. L. Israel\inst{4} \and
D. Lazzati\inst{12} \and
A. Melandri\inst{13,4} \and
L. J. Pellizza\inst{14} \and
P. Romano\inst{2,5} \and
G. Stratta\inst{7} \and
S. D. Vergani\inst{15,16}
}

\offprints{D. Malesani\\\email{malesani@astro.ku.dk}}

\institute{
Dark Cosmology Centre, Niels Bohr Institute, University of Copenhagen, Juliane Maries vej 30, DK-2100 K\o{}benhavn \O, Denmark. \and
INAF, Osservatorio Astronomico di Brera, via E. Bianchi 46, I-23807, Merate (LC), Italy. \and
Universit\`a dell'Insubria, Dipartimento di Fisica e Matematica, via Valleggio 11, I-22100 Como, Italy. \and
INAF, Osservatorio Astronomico di Roma, via Frascati 33, I-00040, Monteporzio Catone (Roma), Italy. \and
Universit\`a degli Studi di Milano-Bicocca, Dipartimento di Fisica, piazza delle Scienze 3, I-20126 Milano, Italy. \and
European Space Astronomy Centre, European Space Agency (ESA), Box 78, 28691 Villanueva de la Ca\~nada, Madrid, Spain. \and
ASI Science Data Center, via G. Galilei, I-00044 Frascati (Roma), Italy. \and
INAF, Osservatorio Astrofisico di Arcetri, largo E. Fermi 5, I-50125, Firenze, Italy. \and
Kavli Institute for Theoretical Physics, University of California, Santa Barbara, CA 93106, USA. \and
Laboratoire Astroparticule et Cosmologie, 10 rue A. Domon et L. Duquet, F-75205 Paris Cedex 13, France. \and
Service d'Astrophysique, DSM/DAPNIA/SAp, CEA-Saclay, F-91191, Gif-sur-Yvette, France. \and
JILA, University of Colorado, Boulder, CO 80309-0440, USA. \and
Astrophysics Research Institute, Liverpool JMU, Twelve Quays House, Egerton Wharfs, Birkenhead, CH41 1LD, U.K. \and
Instituto de Astronom\'{\i}a y F\'{\i}sica del Espacio (CONICET/UBA), Casilla de Correos 67, Suc. 28 (1428) Buenos Aires, Argentina. \and
Dunsink Observatory, DIAS, Dunsink lane, Dublin 15, Ireland. \and
School of Physical Sciences and NCPST, Dublin City University, Dublin 9, Ireland.
}

\date{Received / Accepted}

\titlerunning{The afterglow of GRB\,050724}
\authorrunning{Malesani et al.}

\abstract
{New information on short/hard gamma-ray bursts (GRBs) is being gathered
thanks to the discovery of their optical and X-ray afterglows. However, some
key aspects are still poorly understood, including the collimation level of the
outflow, the duration of the central engine activity, and the properties of the
progenitor systems.}
{We want to constrain the physical properties of the short GRB\,050724 and of
its host galaxy, and make some inferences on the global short GRB population.}
{We present optical observations of the afterglow of GRB\,050724 and of its
host galaxy, significantly expanding the existing dataset for this event. We
compare our results with models, complementing them with available measurements
from the literature. We study the afterglow light curve and spectrum including
X-ray data. We also present observations of the host galaxy.}
{The observed optical emission was likely related to the large flare observed
in the X-ray light curve. The apparent steep decay was therefore not due to the
jet effect. Available data are indeed consistent with low collimation, in turn
implying a large energy release, comparable to that of long GRBs. The flare
properties also constrain the internal shock mechanism, requiring a large
Lorentz factor contrast between the colliding shells. This implies that the
central engine was active at late times, rather than ejecting all shells
simultaneously. The host galaxy has red colors and no ongoing star formation,
consistent with previous findings on this GRB. However, it is not a pure
elliptical, and has some faint spiral structure.}
{GRB\,050724 provides the most compelling case for association between a short
burst and a galaxy with old stellar population. It thus plays a pivotal role in
constraining progenitors models, which should allow for long delays between
birth and explosion.}


\maketitle

\begin{figure*}
\includegraphics[width=\textwidth]{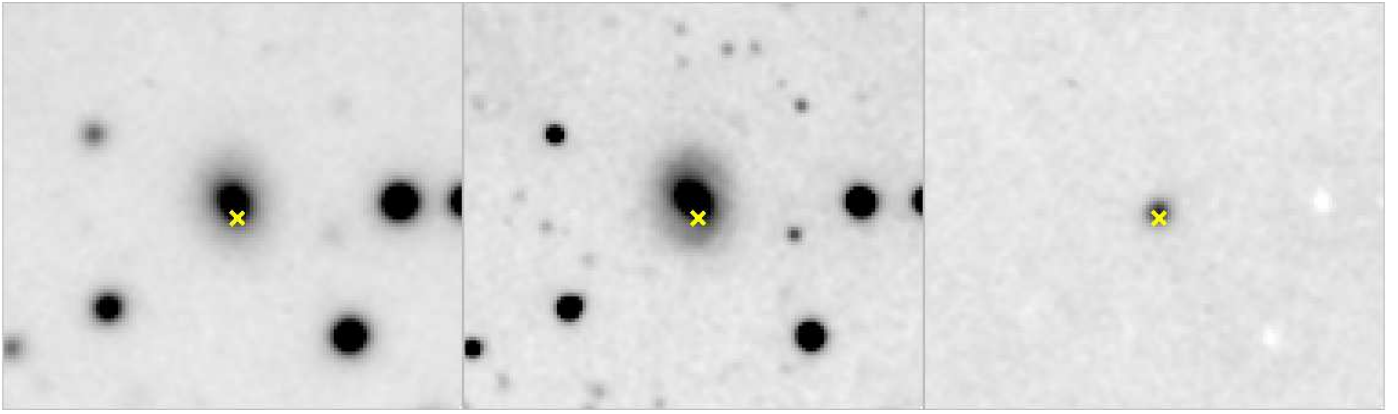}
\caption{$R$-band images of the field of GRB\,050724 about 0.5 (left) and 5.6
(middle) days after the burst. The right panel gives the result of the
subtraction, showing the optical afterglow in the first epoch. Each box is
$26\arcsec \times 22\arcsec$ wide. North is up and east is left. The cross
marks the position of the radio afterglow \citep{Berg05}.\label{fig:subtract}}
\end{figure*}

\begin{table*}
\caption{Log of the observations of GRB\,050724. All measurements were carried
out using the FORS\,1 instrument of the ESO VLT UT2 (Kueyen). Upper limits are 
at the $3\sigma$ confidence level. The epochs marked as ``Reference'' were used
as late-time templates for the subtraction process, and the afterglow
brightness cannot be computed from these images. The magnitudes are not
corrected for Galactic or intrinsic extinction.\label{tb:obslog}}
\centering\begin{tabular}{lllllll} \hline \hline
Mean time $t$     & $t-t_0$  & Filter/ & Exposure time & Seeing   & Airmass & Magnitude      \\
(UT)              & (day)    & grism   & (s)           &(\arcsec) &         &                \\ \hline
2005 Jul 25.01581 & 0.49210  & $I$     & 3$\times$180  & 1.0      & 1.01    & 21.18$\pm$0.03 \\
2005 Jul 25.97533 & 1.45162  & $I$     & 3$\times$180  & 0.8      & 1.06    & 23.22$\pm$0.12 \\
2005 Jul 27.98413 & 3.46042  & $I$     & 3$\times$180  & 0.8      & 1.04    & 25.53$\pm$0.33 \\
2005 Jul 30.11632 & 5.59261  & $I$     & 4$\times$180  & 0.5      & 1.14    & Reference      \\ \hline
2005 Jul 25.00747 & 0.48376  & $R$     & 3$\times$180  & 1.1      & 1.02    & 21.85$\pm$0.04 \\
2005 Jul 25.98390 & 1.46019  & $R$     & 3$\times$180  & 1.0      & 1.05    & 23.66$\pm$0.09 \\
2005 Jul 26.96903 & 2.44532  & $R$     & 1$\times$60   & 0.8      & 1.07    & $>24.4$        \\
2005 Jul 27.97569 & 3.45198  & $R$     & 3$\times$180  & 0.8      & 1.05    & $>24.8$        \\
2005 Jul 30.10470 & 5.58099  & $R$     & 4$\times$180  & 0.5      & 1.11    & Reference      \\
2005 Aug 25.98876 & 32.46505 & $R$     & 8$\times$120  & 0.7      & 1.01    & $> 25.7$       \\ \hline
2005 Jul 24.99906 & 0.47535  & $V$     & 4$\times$120  & 0.9      & 1.03    & 22.49$\pm$0.03 \\
2005 Jul 27.99267 & 3.46896  & $V$     & 3$\times$180  & 0.8      & 1.03    & $> 25.45$      \\
2007 Mar 15.32476 & 598.7975 & $V$     & 6$\times$120  & 0.6      & 1.10    & Reference      \\ \hline
2005 Jul 26.99009 & 2.46638  & 300V    & 3$\times$600  & 1.1      & 1.03    & Spectrum       \\ \hline
\end{tabular}
\end{table*}

\section{Introduction}

Our knowledge of the short/hard class of gamma-ray bursts
\citep[GRBs;][]{Dezalay91,Kouv93} has substantially improved since the launch
of the \textit{Swift} and HETE-2 satellites \citep{Gehe04,Ricker02}. At the
time of writing (2007 April), some 25 events had been accurately localized,
and, for a significant fraction of them, X-ray ($\sim 65$\%), optical ($\sim
30$\%) and radio ($\sim 8$\%) afterglows were detected
\citep{Gehrels05,Villasenor05,Fox05,Hjor05,Cov06,Barth05,Berg05,Soderberg06,Burrows06,LaParola06,Levan06,deUgarte06,Roming06,Berger07}.
This has made possible the identification of their host galaxies (for most of
those with arcsecond localization:
\citealt{Bloom06,Hjorth05b,CT05,Proc06,Goro06,Ferrero06,Berger07}). We refer to
\citet{Naka07} for a recent review on the observational status and its
implications. Despite this progress, the study of short GRB afterglows is still
in its infancy, and only in a few cases are detailed observations available.
Typically, the sampling of the afterglow light curves is poor and broad-band
data are lacking, also due to the intrinsic faintness of these events
\citep{Berger07}. In general, the afterglows of short/hard GRBs have shown an
overall similarity with those of their long-duration brethren, with power-law
decays interrupted by breaks and flares. Basic quantities, however, are still
poorly constrained, such as the true energy release. In fact, the degree of
collimation of their ejecta is still largely unknown, due to the sparse
sampling of afterglow light curves \citep[e.g.][]{Wats06}.

The \textit{Swift} and HETE-2 results have also challenged the standard
division of GRBs into two families based on duration and spectral hardness,
fostering the search of new classification schemes
\citep{Donaghy06,OBWi07,Zha07}. Long-lasting ($\sim 100$~s), soft emission
following short GRBs was revealed, sometimes comprising a major fraction of the
total fluence \citep{Villasenor05,Barth05,NoBe06,Lazzati01}. A possible
extreme example of this behavior is GRB\,060614 \citep{Gehe06,Zha07,Mangano07},
a long-duration GRB with deep limits on any associated supernova
\citep{DeVa06,Gal06,Fynb06}.

GRB\,050724 \citep{Cov05} is one of the most interesting short/hard GRBs
discovered so far. It was the second of this class with an optical and
near-infrared (NIR) counterpart \citep{Berg05,PdA05,Cobb05,Klaas05}, and the
first with detectable radio emission \citep{Cameron05,Berg05}. It is also the
prototype of short/hard GRBs with long-lasting soft emission
\citep{Barth05,Camp06}. The afterglow was found overlaid on a bright ($L \ga
L_*$) galaxy at redshift $z = 0.258$ with very low star formation
\citep[$<0.05\,M_\odot~\mathrm{yr}^{-1}$;][]{Berg05,Proc06} and an old stellar
population \citep[$> 2.6$~Gyr;][]{Goro06}. GRB\,050724 currently is the best
case for association between a GRB and an early-type galaxy.

We present here optical observations of the afterglow and host galaxy of
GRB\,050724. Our data, described in Sect.~\ref{sec:obs}, nearly double the
available dataset for this event. The afterglow and host galaxy are discussed
in Sect.~\ref{sec:afterglow} and \ref{sec:host}, respectively, and we comment
on our results in Sect.~\ref{sec:conc}. Throughout the paper, the decay and
spectral indices $\alpha$ and $\beta$ are defined by $F_\nu(t,\nu) \propto
(t-t_0)^{-\alpha}\nu^{-\beta}$, where $t_0$ is the burst trigger time (2005 Jul
24.52371 UT). We assume a $\Lambda$CDM cosmology with $\Omega_{\rm m} = 0.27$,
$\Omega_\Lambda = 0.73$ and $h_0 = 0.71$ \citep{Spergel03}. At the GRB redshift
($z = 0.258$), the luminosity distance is 1.30~Gpc, the distance modulus is
40.56~mag, and 1\arcsec{} corresponds to 3.97 kpc. All errors are at the
$1\sigma$ confidence level unless stated otherwise.

\section{Observations and data analysis}
\label{sec:obs}

We observed the field of GRB\,050724 with the ESO Very Large Telescope (VLT),
using the FORS1 instrument, starting 0.5~days after the GRB. Imaging in the
$V$, $R$ and $I$ bands was carried out during several of the subsequent nights.
Table~\ref{tb:obslog} provides a summary of our observations. Flux calibration
was achieved by observing the Landolt standard field PG\,1323$-$086 during
several photometric nights. The zeropoint was found to be stable up to $\approx
0.02$ mag. Inside the XRT error circle \citep{Barth05}, the bright galaxy first
noted by \citet{Bloom05} is clearly visible. From our late-time, best-seeing
images, we measured its magnitudes to be $V = 20.45 \pm 0.01$, $R = 19.47 \pm
0.01$, and $I = 18.59 \pm 0.01$ mag (without any extinction correction). These
values are $\approx 0.1$~mag brighter than those reported by \citet{Goro06},
which may reflect our ability to account for the low-surface brightness regions
of the galaxy, or may simply be due to a calibration mismatch. For reference,
we provide in Table~\ref{tab:std} the magnitudes of a few stars in the GRB
field which we adopted as secondary calibrators. Aperture photometry of the
host galaxy revealed a clear dimming in all filters between the first and
subsequent epochs, providing evidence of the presence of the fading afterglow.
To obtain more accurate results, PSF-matched image subtraction was performed
using the ISIS package \citep{AlLu98}. Late-time images with good seeing were
adopted as templates for galaxy subtraction, yielding a detection of the
afterglow in all filters at several epochs (Fig.~\ref{fig:subtract}). The
afterglow flux was determined by comparison with that of artificial stars of
known magnitude inserted in the original images. We expect little afterglow
contribution in the reference images ($< 10$\%). We explicitly checked this in
the $R$ band, where we adopted two different reference images ($\approx 5.5$
and 32.5~days after the GRB; see Table~\ref{tb:obslog}), and obtained
consistent results. Our final photometry is reported in Table~\ref{tb:obslog},
and supersedes our preliminary report \citep{PdA05}.

From the subtraction images, we could accurately determine the position of the
afterglow, which was located at the coordinates $\mbox{RA} = 16^{\rm h}24^{\rm
m}44\fs38$, $\mbox{Dec} = -27\degr32\arcmin27\farcs1$ (J2000, 0\farcs35 RMS
error, relative to 300 USNO-B1 stars). These compare well with those of
\citet{Berg05}, and are also consistent with the X-ray
\citep{Burrows05,Barth05} and radio \citep{Soderberg05,Berg05} positions. The
afterglow is thus 0\farcs6 off the center of the host galaxy, which corresponds
to 2.6~kpc in projection at $z = 0.258$.

Spectroscopy of the host galaxy was obtained during the night of 2006 Jul 26,
using a slit 1\arcsec{} wide and the 300V grism (7.5~\AA{} resolution),
covering the 3800--9000~\AA{} wavelength range. At that epoch the afterglow was
only marginally contributing to the total light ($< 2$\%). Standard
spectroscopic reduction was performed using IRAF\footnote{IRAF is distributed
by the National Optical Astronomy Observatories, which are operated by the
Association of the Universities for Research in Astronomy, Inc., under
cooperative agreement with the National Science Foundation.}. The spectra were
wavelength- and flux-calibrated by using a He-Ar lamp and observing the
spectroscopic standard star LTT\,6248. Slit losses were corrected for by
matching the measured fluxes to the photometry. A simple rescaling by a factor
of 2.3, independent of the wavelength, was enough to account for the
difference. This correction is consistent with the angular size of the galaxy
(half-light radius of $\approx 1\farcs5$).

\begin{table}
\caption{Magnitudes of reference stars in the field of GRB\,050724.\label{tab:std}}
\begin{tabular*}{\columnwidth}{@{\extracolsep{\fill}}c@{\hfill}c@{\hfill}c@{\hfill}c@{\hfill}l}\hline\hline
RA (J2000)                    & Dec (J2000)                  & $V$             & $R$             & $I$            \\ \hline
$16^{\rm h}24^{\rm m}44\fs12$ & $-27\degr33\arcmin37\farcs2$ & 18.74$\pm$0.02  & 18.18$\pm$0.01  & 17.64$\pm$0.01 \\
$16^{\rm h}24^{\rm m}38\fs73$ & $-27\degr32\arcmin17\farcs9$ & 20.01$\pm$0.01  & 19.45$\pm$0.01  & 18.82$\pm$0.02 \\
$16^{\rm h}24^{\rm m}44\fs74$ & $-27\degr30\arcmin59\farcs9$ & 19.71$\pm$0.01  & 19.14$\pm$0.01  & 18.59$\pm$0.02 \\ \hline
\end{tabular*}
\end{table}

\section{Afterglow properties}
\label{sec:afterglow}

\begin{figure}
  \includegraphics[width=\columnwidth]{lc_revised.epsi}
  \caption{X-ray, optical and radio light curves of the afterglow of
  GRB\,050724. Filled and empty symbols represent measurements from our data
  and from the literature, respectively
  \citep{Berg05,Chester05,Torii05,Cobb05,Klaas05,Pastorello05}. The $V$-, $R$-
  and $K$-band data have been displaced vertically for graphical purposes (see
  legend). No correction for optical extinction has been applied. The dotted
  and dashed lines show the best power-law fits to the optical and X-ray data,
  respectively. The vertical dot-dashed line marks the time at which we
  computed the SED (Fig.~\ref{fig:sed}).\label{fig:lightcurve}}
\end{figure}

A collection from the literature of the optical and near-infrared photometry of
the GRB\,050724 afterglow reveals some discrepancies ($\approx 0.5$~mag) when
comparing simultaneous data. This is not surprising, given the intrinsic
difficulties involved in the image subtraction process, especially critical
given the brightness of the host galaxy. Furthermore, several data taken from
the GCN circulars\footnote{\texttt{http://gcn.gsfc.nasa.gov}\,.} might suffer
from a preliminary photometric calibration. Last, some of these measurements
were computed adopting, as reference images for the subtraction, exposures
relatively close in time to the GRB, and possibly contaminated by residual
afterglow light. We note, however, that our first $I$-band point ($t-t_0
\approx 0.5$~days) is fully consistent with a contemporaneous measurement by
\citet{Berg05}.

Figure~\ref{fig:lightcurve} shows our measurements (filled symbols), together
with those available from the literature (empty symbols). X-ray data from
\textit{Swift} (BAT and XRT) and \textit{Chandra} (ACIS-S) were taken from
\citet{Camp06} and \citet{Grup06}, respectively. Radio
data\footnote{\texttt{http://www.aoc.nrao.edu/$\sim$dfrail/allgrb\_table.shtml}\,.}
are from \citet{Berg05}. For self-consistency, we initially performed the fits
using our data only. The afterglow is detected up to 3.5~days after the GRB.
Assuming a power-law behavior, the decay slopes are $\alpha_I = 1.74 \pm 0.09$,
$\alpha_R = 1.51 \pm 0.09$, and $\alpha_V > 1.38$ in the $I$, $R$ and $V$
bands, respectively. Fitting the whole dataset together, we obtain $\alpha_{\rm
opt} = 1.64 \pm 0.06$ ($\chi^2_{\rm r} = 3.2/2$). \citet{Berg05} found a
steeper slope $\alpha_K > 1.9$ in the $K$ band. It is unclear whether this
discrepancy has some physical significance, but we caution that few points are
available, and that the light curve might not be represented by a pure power
law (see below). Apart from the precise value, the decay index is quite steep,
and this led \citet{Berg05} to propose that the light curve had a break before
the beginning of their observations ($\approx 0.5$~days after the GRB). The
early UVOT $V$-band upper limit at $t-t_0 \approx 20$~min \citep{Chester05},
coupled with our measurements, also implies a flatter decay at $t - t_0 \la
0.5$~days (Fig.~\ref{fig:lightcurve}). If interpreted as a jet break, such a
limit on the break time would imply a jet half-opening angle $\vartheta_{\rm
jet} \la 8.5\degr$ \citep{Berg05}.

An inspection of the X-ray data (Fig.~\ref{fig:lightcurve}), however, suggests
a different possibility. Similar to that observed in many long/soft GRBs
\citep{Taglia05,Nous06}, the X-ray light curve shows a steep decay
($\alpha_{\rm X} \approx 3.6$) which becomes flatter at $\sim 800$~s
\citep{Camp06}. From this time on, the decay can be described by a steadily
declining component with flaring activity superimposed. Most noticeable is the
large flare peaking at $\sim 50$~ks (observer frame time). If this flare is
interpreted as being due to a different component (e.g., late activity from the
central engine:
\citealt{FanWei05,Zhang06,Perna06,Proga06,Dai06,Lazzati07,Chin07}), the forward
shock emission does not show any break until at least $\sim 3$~weeks after the
GRB. This would imply a low degree of collimation, with $\vartheta_{\rm jet}
\ga 25\degr$ \citep{Grup06}.

\begin{figure}
\includegraphics[width=\columnwidth]{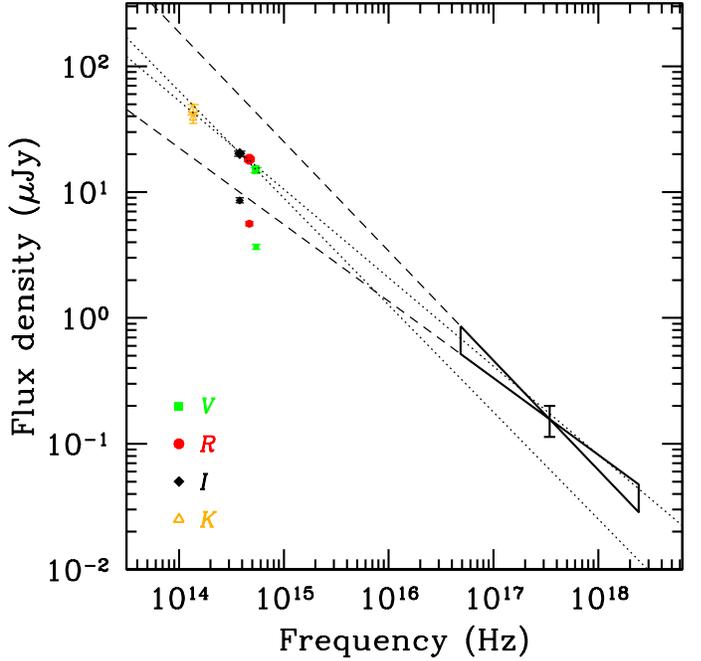}
\caption{Optical/X-ray SED at $t - t_0 = 41.8$~ks. Data were taken from our VLT
  images ($V$, $R$ and $I$ bands) and from \citet[][$K$ band]{Berg05}. Small
  symbols show the observed fluxes, while the large ones indicate the values
  corrected for extinction in the Milky Way, assuming $E(B-V) = 0.46$~mag. The
  dotted and dashed lines show the extrapolation of the optical and X-ray
  spectra, respectively.\label{fig:sed}}
\end{figure}

The discrepancy in the determination of the jet angle may be solved by
considering that all the optical data were taken simultaneously with the large
flare peaking at $t-t_0 \approx 50$~ks, which likely contributed in the optical
band as well. The possible detection of a rising light curve ($F \propto
t^{1.7}$) in the $I$ band between 43 and 51~ks \citep{Berg05} provides some
support for this hypothesis. To further test this possibility, we built the
spectral energy distribution (SED) of the counterpart at 41.8~ks after the
burst. This epoch was chosen because multiband data are available and because
the afterglow was detected with high signal to noise (S/N) ratio. The major
uncertainty is actually the level of the Galactic extinction, which is quite
large towards this region of the sky ($l = 350\degr$, $b = +15\degr$).
Furthermore, as pointed out by \citet{Vaughan06}, this line of sight passes
close to the Ophiuchus molecular cloud complex, making the extinction curve and
the dust-to-gas ratio uncertain. The maps by \citet{Sche98} provide $E(B-V) =
0.61$~mag, but they are known to be scarcely accurate in highly extinguished
regions. \citet{Dutr03} have shown that, in this $E(B-V)$ range, the actual
extinction is lower by a factor of 0.75, with a scatter of $< 20$\%. We
therefore assume $E(B-V) = 0.46$~mag, bearing in mind the uncertainty
associated with this value. For the X-ray spectral slope, we adopted
$\beta_{\rm X} = 0.74 \pm 0.13$ (90\% uncertainty), the average value reported
by \citet{Camp06} over the flare interval (which is consistent with the
measurement by \citealt{Grup06}).

The resulting SED is shown in Fig.~\ref{fig:sed}. With the assumed extinction,
and using our $VRI$ measurements together with the nearly simultaneous $K$-band
detection by \citet{Berg05}, the optical spectral slope is $\beta_{\rm opt} =
0.78 \pm 0.07$. No extinction is assumed close to the burst explosion site, as
expected for a galaxy with an old stellar population and in agreement with
existing estimates \citep{Goro06}. Overall, the SED is consistent with a single
power law extending from the optical to the X-ray ranges, as suggested by the
similarity of $\beta_{\rm opt}$, $\beta_{\rm X}$, and the broad-band spectral
index $\beta_{\rm OX} = 0.72 \pm 0.04$. We note that a perfect match would
require $E(B-V) = 0.49$~mag, which is very similar to the adopted value and
within the scatter of the correction proposed by \citet{Dutr03}. An optical
rebrightening simultaneous with an X-ray flare was also proposed by
\citet{Wats06} for the short GRB\,050709, again removing the need for a break
to explain the steep optical decay. Small-amplitude wiggles have also been
observed in the afterglows of the short GRB\,060121 \citep{deUgarte06} and
GRB\,060313 \citep{Roming06}.

If the optical and X-ray data belong to the same component, we would expect the
same temporal behavior in the two bands, while the decay in the optical is
slower than in the X-rays ($\alpha_{\rm X} = 2.98 \pm 0.15$ during the flare
decline). We note, however, that the optical slope we computed is likely to be
underestimated. In fact, as reported by \citet{Berg05}, the optical flux was
rising at the time of our first observation ($t-t_0 \approx 41.8$~ks), as in
the X-rays. Therefore, since we do not know the optical peak time, and the
light curve is poorly sampled, we can provide only a lower limit to the optical
slope. To estimate the effect of this uncertainty, we took all the available
$I$-band points, including those by \citet{Berg05}, and fitted only those taken
at $t - t_0 > 50$~ks. In this case, we do indeed obtain a steeper value
$\alpha_I = 2.27 \pm 0.14$. Finally, we note that the observed decay rates can
be different in the two bands if the contribution from the underlying forward
shock emission was different, especially at late times.

In Fig.~\ref{fig:lightcurve} we also show the available radio measurements. A
rebrightening is visible in this band too, at a time somehow delayed with
respect to the X-rays. It is not clear whether these two components are
related. \citet{Pana06} explained the radio peak as being due to the passage of
the forward shock injection frequency through the observed band. Similar
behavior was observed in several other afterglows, and the flaring activity at
high energy is not needed to explain the radio light curve.

\subsection{The X-ray flare}

By modeling a smaller data set, \citet{Pana06} suggested that the cooling
frequency was below the optical band at $t-t_0 = 0.5$~days. His analysis,
however, assumed that the optical emission was due to the forward shock.
Furthermore, he assumed a lower extinction $E(B-V) = 0.26$~mag
\citep{BurstHeil82}.

As discussed above, however, our data support a different interpretation,
namely that the observed emission at 0.2--3~days was related to the large flare
apparent in the X-ray light curve. Extensive studies have shown that such
flares cannot be produced in the forward shock, but are the result of late-time
activity of the GRB central engine \citep{Burrows05b,Zhang06,Chin07}, possibly
late internal shocks. Independent of the interpretation of the optical data,
the hard X-ray spectral index (average $\beta \approx 0.74$) suggests that the
peak energy $E_{\rm p}$ was above the XRT band during the flare. The location
of $E_{\rm p}$ can be used to constrain the emission process, under the
hypothesis that the flare was produced by synchrotron radiation in a late
internal shock. Using Eq.~(17) of \citet{ZhangMesz02}, we have
\begin{equation}
  E_{\rm p} \sim 160 \xi L_{52}^{1/2} R_{13}^{-1}~\mathrm{keV},
\end{equation}
where $L = 10^{52}L_{52}$~erg~s$^{-1}$ is the flare luminosity, $R = 10^{13}
R_{13}$~cm is the emission radius, and $\xi$ is a numerical coefficient
dependent on the details of the emission process. By imposing $E_{\rm p} \ga
5$~keV and using the measured isotropic luminosity $L_{\rm flare} = 6 \times
10^{44}$~erg~s$^{-1}$ (0.3--10 keV), we infer a radius $R_{13} \la 0.01\xi$.
For the fireball to be optically thin to Thomson scattering, furthermore,
$R_{13} \ga 1$ is required, and hence $\xi \ga 100$. The parameter $\xi$ is
dependent upon a number of variables, and may be expressed as $\xi = \xi_0
(\Gamma_{12}-1)^\kappa$, where $\xi_0 < 1$, $\Gamma_{12}$ is the relative
Lorentz factor between the colliding shells, and $\kappa = 2$ or $\kappa = 2.5$
depending on the shock parameters \citep{ZhangMesz02}. It is apparent that a
large $\xi$ can be obtained only if $\Gamma_{12} \gg 1$. The data, therefore,
constrain $\Gamma_{12} \approx \Gamma_2/(2\Gamma_1) \ga 10$.

This result has important consequences for the physics of the central engine. A
large $\Gamma_{12}$ implies that the impacting shell was emitted from the
central engine long after the main burst, rather than simultaneously. In fact,
if $\Delta t$ is the time interval between the ejection of two shells, simple
kinematic arguments \citep{Lazzati99} imply
\begin{equation}
  t_{\rm flare} \approx (1+z) \frac{4\Gamma_{12}^2}{4\Gamma_{12}^2-1} \Delta t.
\end{equation}
When $\Gamma_{12} \gg 1$, $\Delta t \approx t_{\rm flare} / (1+z) = 40$~ks.
This result, based on the spectral properties of the flare, agrees with that
inferred by studying flare light curves \citep{Lazzati07}.

\section{The host galaxy}
\label{sec:host}

\begin{figure}
  \includegraphics[width=\columnwidth]{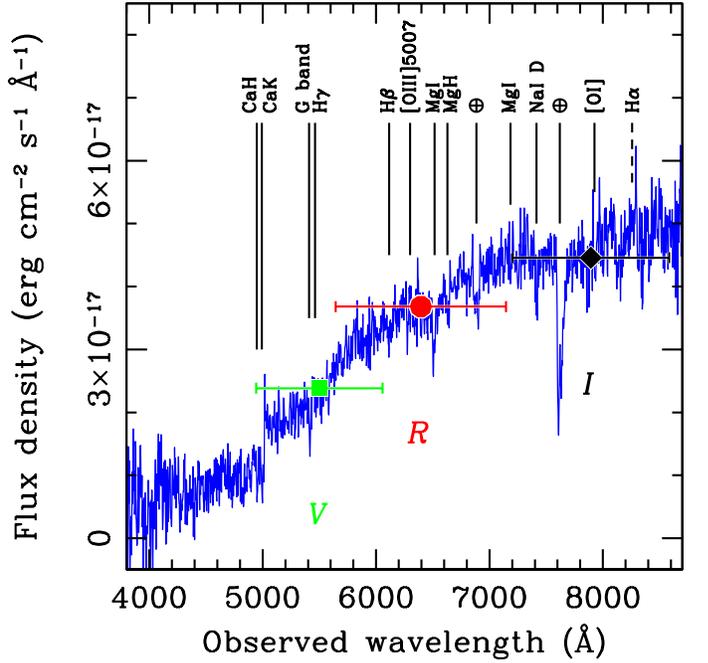}
  \caption{Spectrum of the host galaxy of GRB\,050724, taken on 2005 Jul 26.99
  UT with VLT+FORS1. Telluric lines are indicated by the symbol $\oplus$. The
  spectrum has been  rescaled by a factor of 2.3 to match our photometric
  measurements (dots). In this plot, no extinction correction has been
  applied.\label{fig:spectrum}}
\end{figure}

\begin{table}\centering
\caption{Absorption lines in the spectrum of the host galaxy of GRB\,050724.
Features marked with an asterisk $^*$ have low significance and were not used
for the redshift computation. The line equivalent widths are in the observer
frame.\label{tb:lines}}
\begin{tabular}{lllll} \hline\hline
Line               & $\lambda$ (rest) & $\lambda$ (observed) & Redshift & EW    \\
                   & (\AA)            & (\AA)                &          & (\AA) \\ \hline
Ca K               & 3933.7           & 4950.9               & 0.2586   & 5.4   \\
Ca H               & 3968.5           & 4995.2               & 0.2587   & 4.8   \\
G band             & 4299.6           & 5416.3               & 0.2597   & 4.4   \\
H$\gamma^*$        & 4340.5           & 5469.7               & 0.2602   & 0.8   \\
H$\beta$           & 4861.3           & 6113.0               & 0.2575   & 0.8   \\
$[\ion{O}{III}]^*$ & 5006.8           & 6300.8               & 0.2584   & 0.5   \\
\ion{Mg}{I}        & 5172.7           & 6504.0               & 0.2574   & 3.3   \\
\ion{Mg}{I}        & 5183.7           & 6516.0               & 0.2570   & 1.6   \\
MgH$^*$            & 5269.0           & 6643.5               & 0.2609   & 1.1   \\
MgI                & 5711.1           & 7182.8               & 0.2577   & 0.9   \\
\ion{Na}{I} D      & 5890.9           & 7413.1               & 0.2584   & 1.0   \\
\ion{Na}{I} D      & 5895.9           & 7420.1               & 0.2585   & 1.5   \\
$[\ion{O}{I}]^*$   & 6300.0           & 7931.7               & 0.2590   & 2.2   \\ \hline
\end{tabular}
\end{table}

We secured photometric and spectroscopic observations of the host galaxy of
GRB\,050724 to assess the nature of the GRB progenitor environment. Our
spectrum is shown in Fig.~\ref{fig:spectrum}, and is typical of an evolved
galaxy with an old stellar population. The colors are consistent with those
measured by \citet{Goro06}, which found a best-fit age larger than 2.6~Gyr for
the dominant stellar population. In the spectrum, no emission features are
detected, but from several absorption lines (Table~\ref{tb:lines}) we could
measure a redshift $z = 0.2582 \pm 0.0003$. This is consistent with previous
determinations \citep{Berg05,Proc06}. We also provide an upper limit to the
H$\alpha$ luminosity, $L < 2.8 \times 10^{40}$~erg~s$^{-1}$ (3$\sigma$,
corrected for slit losses and Galactic extinction). Following \citet{Kenn98},
this corresponds to a star formation rate (SFR) $\mbox{} <
0.17~M_\odot$~yr$^{-1}$. The absolute magnitude of the galaxy is $M_B = -21.2$
($L \approx 1.2L_*$ assuming $M^*_B = -21$), computed from the measured
$V$-band flux, so that the SFR per unit luminosity is $<
0.14~M_\odot~\mathrm{yr}^{-1}~L_*^{-1}$. This limit is $\sim 50$ times lower
than the average value found in long-duration GRB hosts, both at low and
intermediate redshift \citep{Sollerman05,Christensen04}. From the available
spectrum, we could also compute a rough estimate of the metallicity, based on
the Mg$_2$ index. Using the theoretical prescription by \citet{Buzzoni92}, and
adopting the age of 2.6~Gyr as determined by \citet{Goro06}, we infer
$\mbox{[Fe/H]} \approx 0.1$. Another estimate was obtained using the G~band,
H$\beta$, Mg$_2$, and \ion{Na}{I} indices and the empirical relations by
\citet{Covino95}. A correction is necessary to account for the age difference
between the GRB host galaxy and the Galactic globular clusters, against which
the empirical relations are calibrated. The inferred metallicity is roughly
solar (with an uncertainty of $\approx 0.2$~dex), which is larger than that
usually observed for long GRB hosts (e.g.
\citealt{Savaglio06,Sollerman05,Stanek06}). We caution, however, that our
determination of the metallicity is appropriate for the stellar component,
while the values inferred for long GRB hosts are relative to the interstellar
medium.

\begin{figure}
  \includegraphics[width=\columnwidth]{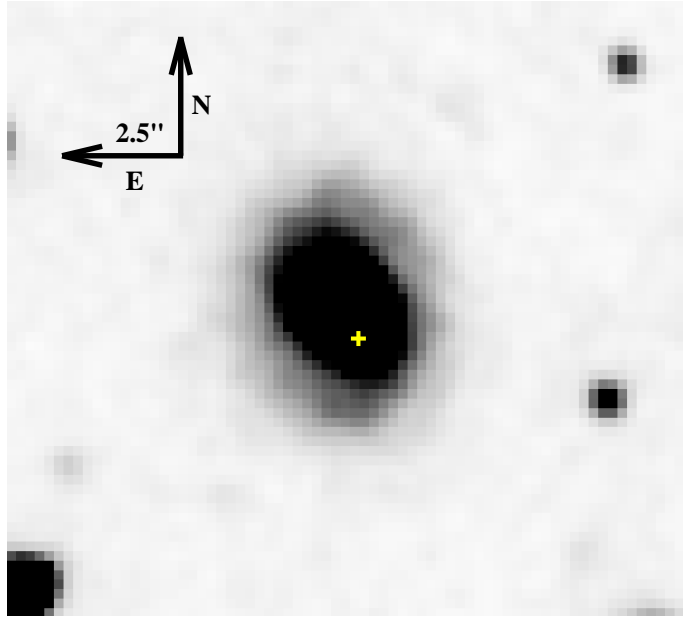}
  \caption{Close-up on the host galaxy of GRB\,050724. The image was taken in
  the $R$ band with VLT+FORS1 on 2005 Jul 30.1 UT. The cross marks the position
  of the optical afterglow. Extended emission is visible towards north-west and
  to the south of the galaxy. The intensity scale is non linear to enhance the
  faint peripheral regions.\label{fg:galaxy}}
\end{figure}

The host galaxy of GRB\,050724 has been morphologically classified as an
elliptical galaxy \citep[e.g][]{Berg05}. Figure~\ref{fg:galaxy} shows an
$R$-band image taken under very good-seeing conditions (0\farcs5) on 2006 July
30.1 UT. The bulge is clearly prominent, but some faint structures are apparent
towards north-west and, to a lesser extent, to the south. These may be due to
weak spiral arms. The galaxy may thus be classified morphologically as an Sa
spiral. To perform a more quantitative analysis, we studied its spatial profile
adopting a two-dimensional fitting approach, applied to our late-time
best-seeing images. To perform the fit we used the image decomposition program
GALFIT \citep{galfit02}, a package  designed to accurately model galaxy
profiles, combining simultaneously an arbitrary number of profiles. The fitting
algorithm constructs a model image, convolves it with the point-spread function
(PSF), and finally compares the result with the data. During the fit, the
reduced $\chi^2$ is minimized using a Levenberg-Marquardt algorithm. The
uncertainties used to calculate the reduced $\chi^2$ as a function of the pixel
position are the Poisson errors, which are generated on the basis of the known
detector characteristics. For each band we constructed the PSF by identifying
in the images $10$ point sources and averaging them. The initial guesses for
the parameters (magnitude, scale length, position angle and minor to major axis
ratio) were obtained by running SExtractor \citep{BertinArnouts96}.

A pure elliptical profile is not a good description for the galaxy morphology.
For $L_*$ galaxies, the surface brightness profile is usually described by the
de Vaucouleurs $r^{1/4}$ law. When applying this model, the residual images
clearly show the spiral arm structure in all the bands, confirming the results
of visual inspection. We allowed for a more general profile function, namely a
Sersic model with free index $n$, but we saw only marginal improvement. The
best-fit index $n = 2.8$ (computed in the $R$ and $I$ bands) is moreover
typical of low-luminosity ellipticals (e.g. \citealt{Caon93}), unlike the host
of GRB\,050724 (which has $L > L_*$). A successful description of the galaxy
was obtained by combining three different components: a de Vaucouleurs profile,
an exponential (disk) function, and a Sersic component. The best-fit values for
the morphological parameters are reported in Table~\ref{tab:galmorph}. The
$V$-band image has a lower S/N ratio, so we list only the results for the $R$
and $I$ bands.

\section{Discussion}
\label{sec:conc}

We have presented an extensive observational campaign characterizing the
afterglow and host galaxy of the short/hard GRB\,050724. We have provided new
data to feed the models and better understand the physical processes occuring
in short GRB fireballs. It is noteworthy that short/hard GRB afterglows share
common properties with those of the long-duration events (see \citealt{Naka07}
for a recent review). For example, independent of the dust extinction, our data
(Fig.~\ref{fig:sed}) show that the optical spectrum has the typical power-law
shape predicted by the synchrotron model.

Based on the existing data, we propose that the steep decay observed in the
optical light curve of the GRB\,050724 afterglow did not result from jetted
emission, but was due to the large X-ray flare which was contributing to the
optical band as well. Such prominent flares are not common, although not
unprecedented (e.g. GRB\,070311: \citealt{Kann07,Guidorzi07}). These large
flares provide a good opportunity to study GRB physics. In particular, if
indeed they are due to late internal activity, they might show measurable
polarization \citep{Fan05} on a timescale easily accessible to large
telescopes. The interpretation of the optical data as belonging to the flare
has important consequences in terms of the energetics of the burst. The low
collimation degree inferred from the X-ray light curve ($\vartheta_{\rm jet}
\ga 25\degr$) implies that the actual explosion energy was not much lower than
the isotropic-equivalent value ($4 \times 10^{50}$~erg; \citealt{Barth05}).
This is comparable to the typical (beaming-corrected) energy release of
long-duration GRBs \citep{Frail01,Ghirla07}. Collimation estimates also affect
the computation of short GRB rates \citep{Nakar06,GuettaPiran06}. To date, the
best evidence of a jetted geometry in short GRBs is provided by the breaks in
the X-ray light curves of GRB\,051221A \citep{Burrows06,Soderberg06} and
GRB\,061201 \citep{Stratta07}, although no late optical and radio data are
available to test the broad-band behavior%
\footnote{The long-duration GRB\,060614, one of the best cases for an
achromatic break in any GRB \citep{Mangano07}, might be related to the short
burst category, but the classification is not conclusive
\citep{Gehe06,Zha07}.}.
In the latter case, a very early break was detected with properties in good
agreement with the expectations of the jet model. Albeit that the redshift of
this GRB is unknown, current limits constrain the beaming-corrected energy to
be less than $10^{49}$~erg, in turn showing that the short GRB luminosity
function is quite broad.

An alternative possibility to explain the different behavior in the optical
and X-ray bands is to assume that the X-ray emission was powered by a
different, long-lived component, as recently suggested for long-duration GRBs
\citep{Pana06b,Willingale06,Uhm07,Genet07,Ghisellini07}. The consistency of the
optical/X-ray SED would in this case be fortuitous.

\begin{table} \caption{Two-dimensional morphological fit parameters of the host
galaxy of GRB\,050724. An asterisk $^*$ denotes a frozen quantity. From top to
bottom: the Sersic index $n$, the scale radius $r_0$, the ratio of the
semi-minor to semi-major axis $b/a$, and the position angle P.A. (measured
counterclockwise relative to the north).\label{tab:galmorph}}
\centering\begin{tabular}{ccccc}\hline\hline
Parameter    & Band & De Vaucouleurs & Exponential & Sersic \\ \hline
$n$          & $I$  & 4$^*$          & 1$^*$       & 1.22   \\
             & $R$  & 4$^*$          & 1$^*$       & 1.12   \\ \hline
$r_0$ $('')$ & $I$  & 1.56           & 1.08        & 1.74   \\
             & $R$  & 1.54           & 1.09        & 1.12   \\ \hline
$b/a$        & $I$  & 0.784          & 0.784       & 0.513  \\
             & $R$  & 0.789          & 0.814       & 0.867  \\ \hline
P.A. (\degr) & $I$  & 23.85          & 23.85       & 40.20  \\
             & $R$  & 24.46          & 23.40       & 21.01  \\ \hline
\end{tabular}
\end{table}

Flares in the light curves of both short and long GRBs have been attributed to
late internal shocks \citep{FanWei05,Zhang06}. There are two variants of this
model. In the first, the colliding shells are ejected together with the main
burst, and impact each other at late times because they a have small velocity
spread. In the second case, the central engine remains active for a long time
and emits fast shells at late times. The X-ray flare of GRB\,050724 was
spectrally hard, with the peak energy above the XRT range. In order to yield
such a high value, the Lorentz factor contrast of the colliding shells had to
be large. This in turn implies that the central engine remained active for a
long time ($\approx 40$~ks), which is not straightforward to achieve in short
GRB models (see \citealt{LeeRR07} for a review).

GRB\,050724 is also remarkable for its association with a galaxy having an old
population. Following the discovery of several short GRB host galaxies, it has
become apparent that this population includes objects with different
properties, and that a significant fraction of short GRBs explode inside
systems with moderate ongoing star formation and relatively young stellar
populations  (\citealt{Fox05,Cov06,Berger07}; P. D'Avanzo et al. 2007, in
preparation). GRB\,050724 provides the most compelling evidence to date that
short GRBs also occur inside galaxies with negligible ongoing star formation.
Using the formulation outlined by \citet{Bloom02}, we estimated the probability
$P$ to find a galaxy brighter than the candidate host\footnote{We applied the
Galactic extinction correction $A_R = 1.2$~mag to the observed value, since
galaxy counts are performed in low-extinction sky regions.} ($R = 18.2$) at an
angular distance less than $0\farcs6$ from the optical afterglow. We found $P$
to be as low as $6.3 \times 10^{-5}$ (see also \citealt{Barth05}), confirming
that the association is not due to a chance superposition. It has been
suggested that other short GRBs are associated with early-type galaxies, most
noticeably GRB\,050509B \citep{Gehrels05,Hjorth05b,Bloom06}. In terms of
progenitors, this implies that either more than one evolutionary channel leads
to the production of short GRBs, or that there is a wide distribution of delay
times between the birth of the progenitor system and the GRB explosion
\citep{Nakar06,GuettaPiran06,ZhengRR06}. The inferred scenario is broadly
consistent with models involving the merging of a binary compact object system
\citep{Eichler89,Belczynski06}. The old age of the host galaxy of GRB\,050724
is also consistent with the lack of detection of a supernova (SN) associated
with this GRB. Our late-time images (32 days after the GRB) constrain the
contribution of a SN at the position of the GRB to be fainter than SN\,1998bw
\citep{Galama98} by at least $\approx 3$~mag in the observed $R$ band. This
corresponds to an absolute magnitude $M_V > -16.2$.

The association of GRB\,050724 with an early-type galaxy is especially
significant given its peculiar prompt light curve shape (a short spike followed
by a long, soft pulse), which would nominally make GRB\,050724 a long-duration
event (formally $T_{90} > 2$~s; \citealt{Barth05}). Its host galaxy is in fact
distinctly different from those of long GRBs (which are typically blue,
subluminous, young and metal-poor; e.g.
\citealt{Djorgovski98,LeFloch03,Fynbo03,Christensen04,Fruchter06}), and is very
unlikely to host young stars akin to the progenitors of long GRBs. This
supports the idea that the duration is not the only parameter relevant to the
classification of bursts, and that some long-lasting GRBs are not associated
with star formation (see also \citealt{Zha07}). Late-time ($\approx 100$~s)
soft emission occurs in a significant fraction of short bursts, $\approx 30$\%
in the \textit{Swift}/HETE-2 sample. This is actually a lower limit, since some
of these events might be confused with long-duration events (e.g. GRB\,050911:
\citealt{Page06}). There seems to be no relation between the host galaxy type
and the presence of the soft component. A well-known case is GRB\,050709
\citep{Villasenor05,Fox05,Cov06}, exploded in a moderately star-forming galaxy,
which also displayed the soft hump. Looking at the present sample, it seems
that bursts with long-lasting emission have more often an optical afterglow
($\approx 70\%$ of the cases) than the overall population, but this is based on
very limited statistics (seven events). Estabilishing the link between the
different kinds of short-duration GRBs will be an important clue to understand
their progenitors.

\begin{acknowledgements}
We acknowledge the ESO staff at Paranal, in particular Jason Spyromillo, and
all the visiting observers who accepted to perform our ToO observations. We
also thank Johan P. U. Fynbo, Jens Hjorth and Eleonora Troja for discussion,
and an anonymous referee for her/his careful reading of the manuscript. The
Dark Cosmology Centre is funded by the Danish National Research Foundation. DM
and MDV acknowledge the Instrument Center for Danish Astrophysics and the
National Science Fundation (grant PHY05-51164), respectively, for financial
support. This work was also funded by ASI grant I/R/039/04 and MIUR grant
2005025417.
\end{acknowledgements}


\begin{thebibliography}{}
\bibitem[Alard \& Lupton(1998)]{AlLu98} Alard, C., \& Lupton, R. H. 1998, \apj, 503, 325
\bibitem[Barthelmy et al.(2005)]{Barth05} Barthelmy, S. D., Chincarini, G., Burrows, D. N., et al. 2005, \nat, 438, 994
\bibitem[Belczynski et al.(2006)]{Belczynski06} Belczynski, K., Perna, R., Bulik, T., et al. 2006, \apj, 648, 1110
\bibitem[Berger et al.(2005)]{Berg05} Berger, E., Price, P. A., Cenko, S. B., et al. 2005, \nat, 438, 988
\bibitem[Berger et al.(2007)]{Berger07} Berger, E., Fox, D. B., Price, P. A., et al. 2007, \apj, 664, 1000
\bibitem[Bertin \& Arnouts(1996)]{BertinArnouts96} Bertin, E., \& Arnout, S. 1996, \aaps, 117, 393
\bibitem[Bloom et al.(2002)]{Bloom02} Bloom, J. S., Kulkarni, S. R., \& Djorgovski, S. G. 2002, \aj, 123, 1111
\bibitem[Bloom et al.(2005)]{Bloom05} Bloom, J. S., Dupree, A., Chen, H.-W., \& Prochaska, J. X. 2005, GCN Circ. 3672
\bibitem[Bloom et al.(2006)]{Bloom06} Bloom, J. S., Prochaska, J. X., Pooley, D., et al. 2006, \apj, 638, 354
\bibitem[Burrows et al.(2005a)]{Burrows05} Burrows, D. N., Grupe, D., Kouveliotou, C., et al. 2005a, GCN Circ. 3697
\bibitem[Burrows et al.(2005b)]{Burrows05b} Burrows, D. N., Romano, P., Falcone, A., et al. 2005b, Science, 309, 1833
\bibitem[Burrows et al.(2006)]{Burrows06} Burrows, D. N., Grupe, D., Capalbi, M., et al. 2006, \apj, 653, 468
\bibitem[Burstein \& Heiles(1982)]{BurstHeil82} Burstein, D., \& Heiles, C. 1982, \aj, 87, 1165
\bibitem[Buzzoni et al.(1992)]{Buzzoni92} Buzzoni, A., Gariboldi, G., \& Mantegazza, L. 1992, \aj, 103, 1814
\bibitem[Cameron \& Frail(2005)]{Cameron05} Cameron, P. B., \& Frail, D. A. 2005, GCN Circ. 3676
\bibitem[Campana et al.(2006)]{Camp06} Campana, S., Tagliaferri, G., Lazzati, D., et al. 2006, A\&A, 454, 113
\bibitem[Caon et al.(1993)]{Caon93} Caon, N., Capaccioli, M., \& D'Onofrio, M. 1993, \mnras, 265, 1013
\bibitem[Castro-Tirado et al.(2005)]{CT05} Castro-Tirado, A. J., de Ugarte Postigo, A., Gorosabel, J., et al. 2005, \aap, 439, L15
\bibitem[Chester et al.(2005)]{Chester05} Chester, M., Covino, S., Schady, P., Roming, P., \& Gehrels, N. 2005, GCN Circ. 3670
\bibitem[Chincarini et al.(2007)]{Chin07} Chincarini, G., Moretti, A., Romano, P., et al. 2007, \apj, in press (astro-ph/0702371)
\bibitem[Christensen et al.(2004)]{Christensen04} Christensen, L., Hjorth, J., \& Gorosabel J. 2004, \aap, 425, 913
\bibitem[Cobb \& Bailyn(2005)]{Cobb05} Cobb, B. E., \& Bailyn, C. D. 2005, GCN Circ. 3694
\bibitem[Covino et al.(1995)]{Covino95} Covino, S., Galletti, S., \& Pasinetti, L. E. 1995, \aap, 303, 79
\bibitem[Covino et al.(2005)]{Cov05} Covino, S., Antonelli, L. A., Romano, P., et al. 2005, GCN Circ. 3665
\bibitem[Covino et al.(2006)]{Cov06} Covino, S., Malesani, D., Israel, G. L., et al. 2006, \aap, 447, L5
\bibitem[Dai et al.(2006)]{Dai06} Dai, Z. G., Wang, X. Y., Wu, X. F., \& Zhang, B. 2006, Science, 311, 1127
\bibitem[D'Avanzo et al.(2005)]{PdA05} D'Avanzo, P., Covino, S., Antonelli, L. A., et al. 2005, GCN Circ. 3690
\bibitem[Della Valle et al.(2006)]{DeVa06} Della Valle, M., Chincarini, G., Panagia, N., et al. 2006, \nat, 444, 1050
\bibitem[de Ugarte Postigo et al.(2006)]{deUgarte06} De Ugarte Postigo, A., Castro-Tirado, A. J., Guziy, S., et al. 2006, \apj, 648, L89
\bibitem[Dezalay et al.(1991)]{Dezalay91} Dezalay, J. P., Barat, C., Talon, R., et al. 1991, in AIP Conf. Proc. 265, Gamma-ray bursts, ed. W. Paciesas \& G. J. Fishman, 304
\bibitem[Djorgovski et al.(1998)]{Djorgovski98} Djorgovski, S. G., Kulkarni, S. R., Bloom, J. S., et al. 1998, \apj, 508, L17
\bibitem[Donaghy et al.(2006)]{Donaghy06} Donaghy, T. Q., Lamb, D. Q., Sakamato, T., et al. 2006, \apj, submitted (astro-ph/0605570)
\bibitem[Dutra et al.(2003)]{Dutr03} Dutra, C. M., Ahumada, A. V., Clari\'a, J. J., Bica, E., \& Barbuy, B. 2003, \aap, 408, 287
\bibitem[Eichler et al.(1989)]{Eichler89} Eichler, D., Livio, M., Piran, T., \& Schramm, D. N. 1989, \nat, 340, 126
\bibitem[Fan \& Wei(2005a)]{FanWei05} Fan, Y. Z., \& Wei, D. M. 2005a, \mnras, 364, L42
\bibitem[Fan et al.(2005b)]{Fan05} Fan, Y. Z., Zhang, B., \& Proga, D. 2005b, \apj, 635, L129
\bibitem[Ferrero et al.(2006)]{Ferrero06} Ferrero, P., Sanchez, S. F., Kann, D. A., et al. 2006, \aj, in press (astro-ph/0610255)
\bibitem[Fox et al.(2005)]{Fox05} Fox, D. B., Frail, D. A., Price, P. A., et al. 2005, \nat, 437, 845
\bibitem[Frail et al.(2001)]{Frail01} Frail, D. A., Kulkarni, S. R., Sari, R., et al. 2001, \apj, 562, L55
\bibitem[Fruchter et al.(2006)]{Fruchter06} Fruchter, A. S., Levan, A. J., Strolger, L., et al. 2006, \nat, 441, 463
\bibitem[Fynbo et al.(2003)]{Fynbo03} Fynbo, J. P. U., Jakobsson, P., M\o{}ller, P., et al. 2003, \aap, 406, L63
\bibitem[Fynbo et al.(2006)]{Fynb06} Fynbo, J. P. U., Watson, D., Th\"one, C. C., et al. 2006, \nat, 444, 1047
\bibitem[Galama et al.(1998)]{Galama98} Galama, T. J., Vreeswijk, P. M., van Paradijs, J., et al. 1998, \nat, 395, 670
\bibitem[Gal-Yam et al.(2006)]{Gal06} Gal-Yam, A., Fox, D. B., Price, P. A., et al. 2006, \nat, 444, 1053
\bibitem[Gehrels et al.(2004)]{Gehe04} Gehrels, N., Chincarini G., Giommi, P., et al. 2004, \apj, 611, 1005
\bibitem[Gehrels et al.(2005)]{Gehrels05} Gehrels, N., Sarazin, C. L., O'Brien, P. T., et al. 2005, \nat, 437, 851
\bibitem[Gehrels et al.(2006)]{Gehe06} Gehrels, N., Norris, J. P., Barthelmy, S., et al. 2006, \nat, 444, 1044
\bibitem[Genet et al.(2007)]{Genet07} Genet, F., Daigne, F., \& Mochkovitch, R. 2007, astro-ph/0701204
\bibitem[Ghisellini et al.(2007)]{Ghisellini07} Ghisellini, G., Ghirlanda, G., Nava, L., \& Firmani, C. 2007, \apj, 658, L75
\bibitem[Ghirlanda et al.(2007)]{Ghirla07} Ghirlanda, G., Nava, L., Ghisellini, G., \& Firmani, C. 2007, \aap, 466, 127
\bibitem[Gorosabel et al.(2006)]{Goro06} Gorosabel, J., Castro-Tirado, A. J., Guziy, S., et al. 2006, \aap, 450, 87
\bibitem[Guetta \& Piran(2006)]{GuettaPiran06} Guetta, D., \& Piran, T. 2006, \aap, 453, 823
\bibitem[Grupe et al.(2006)]{Grup06} Grupe, D., Burrows, D. N., Patel, S. K., et al. 2006, \apj, 653, 462
\bibitem[Guidorzi et al.(2007)]{Guidorzi07} Guidorzi, G., Vergani, S. D., Sazonov, S., et al. 2007, \aap, in press (arXiv:0708.1383)
\bibitem[Hjorth et al.(2005a)]{Hjor05} Hjorth, J., Watson, D., Fynbo, J. P. U., et al. 2005a, \nat, 437, 895
\bibitem[Hjorth et al.(2005b)]{Hjorth05b} Hjorth, J., Sollerman, J., Gorosabel, J., et al. 2005b, \apj, 630, L117
\bibitem[Kann(2007)]{Kann07} Kann, D. A. 2007, GCN Circ. 6209
\bibitem[Kennicutt(1998)]{Kenn98} Kennicutt, R. C. 1998, \araa, 36, 189
\bibitem[Kouveliotou et al.(1993)]{Kouv93} Kouveliotou, C., Meegan, C. A., Fishman, G. J., et al. 1993, \apj, 541, L101
\bibitem[La Parola et al.(2006)]{LaParola06} La Parola, V., Mangano, V., Fox., D., et al. 2006, \aap, 454, 753
\bibitem[Lazzati et al.(1999)]{Lazzati99} Lazzati, D., Ghisellini, G., \& Celotti, A. 1999, \mnras, 309, L13
\bibitem[Lazzati et al.(2001)]{Lazzati01} Lazzati, D., Ramirez-Ruiz, E., \& Ghisellini, G. 2001, \aap, 379, L39
\bibitem[Lazzati \& Perna(2007)]{Lazzati07} Lazzati, D., \& Perna, R. 2007, \mnras, 375, L46
\bibitem[Lee \& Ramirez-Ruiz(2007)]{LeeRR07} Lee, W. H., \& Ramirez-Ruiz, E. 2007, New J. Phys., 9, 17
\bibitem[Le Floc'h et al.(2003)]{LeFloch03} Le Floc'h, E., Duc, P. A., Mirabel, I. F., et al. 2003, \aap, 400, 499
\bibitem[Levan et al.(2006)]{Levan06} Levan, A. J., Tanvir, N. R., Fruchter, A. S., et al. 2006, \apj, 648, L9
\bibitem[Mangano et al.(2007)]{Mangano07} Mangano, V., Holland, S. T., Malesani, D., et al. 2007, \aap, 470, 105
\bibitem[Nakar et al.(2006)]{Nakar06} Nakar, E., Gal-Yam, A., \& Fox, D. B. 2006, \apj, 650, 281
\bibitem[Nakar(2007)]{Naka07} Nakar, E. 2007, Phys. Rep., 442, 166
\bibitem[Norris \& Bonnell(2006)]{NoBe06} Norris, J. P., \& Bonnell, J. T. 2006, \apj, 643, 266
\bibitem[Nousek et al.(2006)]{Nous06} Nousek, J. A., Kouveliotou, C., Grupe, D., et al. 2006, \apj, 642, 389
\bibitem[O'Brien \& Willingale(2007)]{OBWi07} O'Brien, P. T., \& Willingale, R. 2007, Phil. Trans. R. Soc. A, 365, 1179
\bibitem[Page et al.(2006)]{Page06} Page, K. L., King, A. R., Levan, A. J., et al. 2006, \apj, 637, L13
\bibitem[Panaitescu(2006a)]{Pana06} Panaitescu, A. 2006a, \mnras, 367, L42
\bibitem[Panaitescu et al.(2006b)]{Pana06b} Panaitescu, A., M\'esz\'aros, P., Burrows, D., et al. 2006b, \mnras, 369, 2059
\bibitem[Pastorello et al.(2005)]{Pastorello05} Pastorello, A., Kawabata, K., Pian, E., et al. 2005, GCN Circ. 3892
\bibitem[Peng et al.(2002)]{galfit02} Peng, C. Y., Ho, L. C., Impey, C. D., \& Rix, H. 2002, \aj, 124, 266
\bibitem[Perna et al.(2006)]{Perna06} Perna, R., Armitage, P. J., \& Zhang, B. 2006, \apj, 636, L29 
\bibitem[Prochaska et al.(2006)]{Proc06} Prochaska, J. X., Bloom, J. S., Chen H.-W., et al. 2006, \apj, 642, 989
\bibitem[Proga et al.(2006)]{Proga06} Proga, D., \& Zhang, B. 2006, \mnras, 370, L61
\bibitem[Ricker et al.(2002)]{Ricker02} Ricker, G. R., Atteia, J.-L., Crew, G. B., et al. 2002, in AIP Conf. Proc. 662, Gamma-Ray Burst and Afterglow Astronomy 2001: A Workshop Celebrating the First Year of the HETE Mission, ed. G. R. Ricker \& R. K. Vanderspek, 3
\bibitem[Roming et al.(2006)]{Roming06} Roming, P. W. A., Vanden Berk, D., Pal'shin, V., et al. 2006, \apj, 651, 985
\bibitem[Savaglio(2006)]{Savaglio06} Savaglio, S. 2006, New J. Phys., 8, 195
\bibitem[Schlegel et al.(1998)]{Sche98} Schlegel, D. J., Finkbeiner, D. P., \& Davis, M. 1998, \apj, 500, 525
\bibitem[Soderberg(2005)]{Soderberg05} Soderberg, A. M. 2005, GCN Circ. 3696
\bibitem[Soderberg et al.(2006)]{Soderberg06} Soderberg, A. M., Berger, E., Kalsiwal, M., et al. 2006, \apj, 650, 261
\bibitem[Sollerman et al.(2005)]{Sollerman05} Sollerman, J., \"Ostlin, G., Fynbo, J. P. U., et al. 2005, \na, 11, 103
\bibitem[Spergel et al.(2003)]{Spergel03} Spergel, D. N., Verde, L., Peiris, H., et al. 2003, \apjs, 148, 175
\bibitem[Stanek et al.(2006)]{Stanek06} Stanek, K. Z., Gnedin, O. Y., Beacom, J. F., et al. 2006, Acta Astron., 56, 333
\bibitem[Stratta et al.(2007)]{Stratta07} Stratta, G., D'Avanzo, P., Piranomonte, S., et al. 2007, \aap, submitted
\bibitem[Tagliaferri et al.(2005)]{Taglia05} Tagliaferri, G., Goad, M., Chincarini, G., et al. 2005, \nat, 436, 985
\bibitem[Torii(2005)]{Torii05} Torii, K. 2005, GCN Circ. 3674
\bibitem[Uhm \& Beloborodov(2007)]{Uhm07} Uhm, Z. L., \& Beloborodov, A. M. 2007, \apj, 665, L93
\bibitem[Vaughan et al.(2006)]{Vaughan06} Vaughan, S., Willingale, R., Romano, P., et al. 2006, \apj, 639, 323
\bibitem[Villasenor et al.(2005)]{Villasenor05} Villasenor, J. S., Lamb, D. Q., Ricker, G. R., et al. 2005, \nat, 437, 855
\bibitem[Watson et al.(2006)]{Wats06} Watson, D., Hjorth, J., Jakobsson, P., et al. 2006, \aap, 454, L123
\bibitem[Wiersema et al.(2005)]{Klaas05} Wiersema, K., Rol, E., Starling, R., et al. 2005, GCN Circ. 3699
\bibitem[Willingale et al.(2007)]{Willingale06} Willingale, R., O'Brien, P. T., Osborne, J. P., et al. 2007, \apj, 662, 1093
\bibitem[Zhang \& M\'esz\'aros(2002)]{ZhangMesz02} Zhang, B., \& M\'esz\'aros, P. 2002, \apj, 581, 1236
\bibitem[Zhang et al.(2006)]{Zhang06} Zhang, B., Fan, Y.-Z., Dyks, J., et al. 2006, \apj, 642, 354
\bibitem[Zhang et al.(2007)]{Zha07} Zhang, B., Zhang, B.-B., Liang, E.-W., et al. 2007, \apj, 655, L25
\bibitem[Zheng \& Ramirez-Ruiz(2007)]{ZhengRR06} Zheng, Z., \& Ramirez-Ruiz, E. 2007, \apj, 665, 1220

\end{thebibliography}
\end{document}